\documentstyle[a4,11pt,epsf,twoside]{article}

\expandafter\ifx\csname mathrm\endcsname\relax\def\mathrm#1{{\rm #1}}\fi

\newcommand{\born}{{\mathrm{Born}}}
\newcommand{\ISR}{{\mathrm{ISR}}}
\def\bfi{\begin{figure}}
\def\efi{\end{figure}}

\def\reffi#1{\mbox{Figure~\ref{#1}}}

\def\refta#1{\mbox{Table~\ref{#1}}}

\def\citere#1{\mbox{Ref.~\cite{#1}}}
\def\citeres#1{\mbox{Refs.~\cite{#1}}}

\newcommand{\TeV}{\unskip\,\mathrm{TeV}}
\newcommand{\GeV}{\unskip\,\mathrm{GeV}}

\newcommand{\fba}{\unskip\,\mathrm{fb}}

\def\mathswitchr#1{\relax\ifmmode{\mathrm{#1}}\else$\mathrm{#1}$\fi}

\newcommand{\PW}{\mathswitchr W}

\newcommand{\PZ}{\mathswitchr Z}

\newcommand{\PH}{\mathswitchr H}

\newcommand{\Pe}{\mathswitchr e}

\newcommand{\Pd}{\mathswitchr d}

\newcommand{\Pu}{\mathswitchr u}

\newcommand{\Pb}{\mathswitchr b}

\newcommand{\Pt}{\mathswitchr t}

\newcommand{\Pep}{\mathswitchr {e^+}}
\newcommand{\Pem}{\mathswitchr {e^-}}

\def\mathswitch#1{\relax\ifmmode#1\else$#1$\fi}

\newcommand{\MH}{\mathswitch {M_\PH}}

\begin{document}
\thispagestyle{empty}
\null
\mbox{} \hfill MPI-PhT/2002-56\\
\strut\hfill KA-TP-16-2002 \\
\strut\hfill hep-ph/0210168 
\vskip .5cm
\vfill
\begin{center}
{\large \bf\boldmath{SIX-FERMION PRODUCTION AT $\Pep\Pem$ COLLIDERS}%
\footnote{To appear in the proceedings of the {\it International Workshop 
on Linear Colliders}, August 26--30, 2002, Jeju Island, Korea.}
\par} \vskip 2em
\vspace{.0cm}
{\large
{\sc Stefan Dittmaier$^1$ and Markus Roth$^{2}$ } } 
\\[.5cm]
$^1$ {\it Max-Planck-Institut f\"ur Physik (Werner-Heisenberg-Institut)\\
D-80805 M\"unchen, Germany}
\\[0.3cm]
$^2$ {\it Institut f\"ur Theoretische Physik, Universit\"at Karlsruhe\\
D-76131 Karlsruhe, Germany}
\par 
\end{center}\par
\vskip 2.0cm {\bf Abstract:} \par 
{
The class of six-fermion production processes at $\Pep\Pem$ colliders
comprises very interesting particle reactions, such as the production of
top-quark pairs and of Higgs bosons in the intermediate Higgs mass range,
the scattering of massive gauge bosons, and triple gauge-boson
production. The Monte Carlo event generator {\sc Lusifer} is designed
for the analysis of such processes. A few illustrating results obtained
with {\sc Lusifer} are discussed.
\par
\vskip 2cm
\vfill
\noindent
October 2002   
\null
\setcounter{page}{0}

\clearpage

\setcounter{page}{1}
\title{SIX-FERMION PRODUCTION AT $\Pep\Pem$ COLLIDERS} 
\author{STEFAN DITTMAIER$^1$%
	\thanks{e-mail address: dittmair@mppmu.mpg.de} 
	 {} and MARKUS ROTH$^2$
\\
\\
        $^1$ {\it Max-Planck-Institut f\"ur Physik (Werner-Heisenberg-Institut)}
\\
	     {\it D-80805 M\"unchen, Germany}
\\
\\
        $^2$ {\it Institut f\"ur Theoretische Physik, Universit\"at Karlsruhe}
\\
             {\it D-76131 Karlsruhe, Germany}
}
\date{}
\maketitle
\begin{abstract}
The class of six-fermion production processes at $\Pep\Pem$ colliders
comprises very interesting particle reactions, such as the production of
top-quark pairs and of Higgs bosons in the intermediate Higgs mass range,
the scattering of massive gauge bosons, and triple gauge-boson
production. The Monte Carlo event generator {\sc Lusifer} is designed
for the analysis of such processes. A few illustrating results obtained
with {\sc Lusifer} are discussed.
\end{abstract}

\section{Introduction}

The Monte Carlo event generator {\sc Lusifer} in its first version,
which is described in \citere{Dittmaier:2002ap} in detail, deals with all 
processes $\Pep\Pem\to 6\,$fermions at tree level in the Standard Model.
The predictions are based on the full set of Feynman diagrams,
the number of which is typically of the order of $10^2$--$10^4$.
Fermions other than top quarks, which are not allowed as external fermions, 
are taken to be massless, and
polarization is fully supported. The helicity amplitudes are
generically calculated with the spinor method of \citere{Dittmaier:1999nn}.
The phase-space integration is based on the
multi-channel Monte Carlo integration technique \cite{Berends:gf},
improved by adaptive weight optimization \cite{Kleiss:qy}.
Channels and appropriate mappings are provided for each 
individual diagram in a generic way. 
More details on the phase-space parametrizations can be found in
\citeres{By73,Denner:1999gp}.
Owing to the potentially large number of Feynman diagrams
per final state, an efficient generic approach has been crucial,
in order to gain an acceptable speed and stability of the program.
Initial-state radiation is included at the leading logarithmic
level employing the structure-function approach (see e.g.\ the 
appendix of \citere{Beenakker:1996kt}).

In the following we collect a few illustrating results of 
\citere{Dittmaier:2002ap} that have been obtained
with {\sc Lusifer}.
In some cases, the tuned comparison with the combination of the
{\sc Whizard} \cite{Kilian:2001qz} and {\sc Madgraph} \cite{Stelzer:1994ta}
packages is included in the discussion.
Specifically, we focus on top-quark pair production,
the production of Higgs bosons in the intermediate Higgs mass range,
and the scattering of massive gauge bosons. 
The precise input for the used parameters and phase-space cuts, as well as
much more results, can be found in \citere{Dittmaier:2002ap}.
We refer to the literature
for further discussions of top-quark pair production 
\cite{Yuasa:1997fa,Accomando:1997gu,Accomando:1997gj},
Higgs-boson production \cite{Accomando:1997gj,Montagna:1997dc},
vector-boson scattering \cite{Gangemi:2000sk},
and triple gauge-boson production \cite{Accomando:1997gu,Accomando:1997gj},
which are also based on full $\Pep\Pem\to 6f$ matrix elements.

\section{Results On Top-Quark Pair Production}

In \refta{tab:bfs} we collect some results on cross sections 
that receive contributions from top-quark pair production, 
$\Pep\Pem\to\Pt\bar\Pt\to 6f$. 
\begin{table}[b]
\centerline{
\begin{tabular}{|c||r@{}l|r@{}l||r@{}l|r@{}l|}
\hline
& \multicolumn{4}{c||}{\sc Lusifer} & 
\multicolumn{4}{c|}{\sc Whizard \& Madgraph}
\\
\cline{2-9}
$\Pep\Pem\to$ & 
\multicolumn{2}{c|}{$\sigma_{\born}[\fba]$} &
\multicolumn{2}{c||}{$\sigma_{\born+\ISR}[\fba]$} &
\multicolumn{2}{c|}{$\sigma_{\born}[\fba]$} &
\multicolumn{2}{c|}{$\sigma_{\born+\ISR}[\fba]$}
\\ \hline\hline
$\mu^- \bar\nu_\mu \nu_\mu \mu^+ \Pb \bar\Pb$  & 5.&8091(49) & 5.&5887(36) 
 & 5.&8102(26) & 5.&5978(30)
\\ \hline
$\mu^- \bar\nu_\mu \nu_\tau \tau^+ \Pb \bar\Pb$  & 5.&7998(36) & 5.&5840(40)
 & 5.&7962(26) & 5.&5893(29)
\\ \hline
$\Pem \bar\nu_\Pe \nu_\mu \mu^+ \Pb \bar\Pb$ & 5.&8188(45) & 5.&6042(38)
 & 5.&8266(27) & 5.&6071(30)
\\ \hline
$\Pem \bar\nu_\Pe \nu_\Pe \Pep \Pb \bar\Pb$ & 5.&8530(68) & 5.&6465(70)
 & 5.&8751(30) & 5.&6508(36)
\\ \hline
$\mu^- \bar\nu_\mu \Pu \bar\Pd \Pb \bar\Pb$ & 17.&095(11) & 16.&4538(98) 
 & 17.&1025(80) & 16.&4627(87)
\\ \hline
$\Pem \bar\nu_\Pe \Pu \bar\Pd \Pb \bar\Pb$  & 17.&187(21)  & 16.&511(12)
 & 17.&1480(82) & 16.&5288(92)
\\ \hline
\end{tabular} }
\caption{Cross sections (without gluon-exchange diagrams)
for top-quark pair production at $\sqrt{s}=500\GeV$}
\label{tab:bfs}
\end{table}
The difference between the cross sections
with two and four quarks in the final states roughly reflects the colour
factor 3
between leptonically and hadronically decaying W~bosons that have been
produced in $\Pt\to\Pb\PW^+$. 
The cross sections are all strongly dominated by the signal diagrams
for resonant $\Pt\bar\Pt$ production, which are identical for all
considered final states. Differences are entirely due to so-called
background diagrams, the size of which is, however, very sensitive
to the angular separation cut between outgoing $\Pe^\pm$ and the beams.
The inclusion of gluon-exchange diagrams would not influence the
integrated cross section significantly. The numbers show that ISR
reduces the cross sections at the level of $\sim 4\%$ at a
centre-of mass (CM) energy of $\sqrt{s}=500\GeV$.
Finally, the comparison of the {\sc Lusifer} and {\sc Whizard \& Madgraph} 
results reveals good agreement.

\section{Results On Higgs-Boson Production}

In \reffi{fig:higgs-nn4q} we show the invariant-mass $(M_{4q})$
and production angular $(\theta_{4q})$
distributions for the four-quark system (including all $4q$
configurations of the first two generations) of the reactions
$\Pep\Pem\to(\nu_\mu\bar\nu_\mu/\nu_\Pe\bar\nu_\Pe)+4q$.
\begin{figure}[p]
\setlength{\unitlength}{1cm}
\centerline{
\begin{picture}(13.5,12.5)
\put(-5.4,- 6.4){\includegraphics{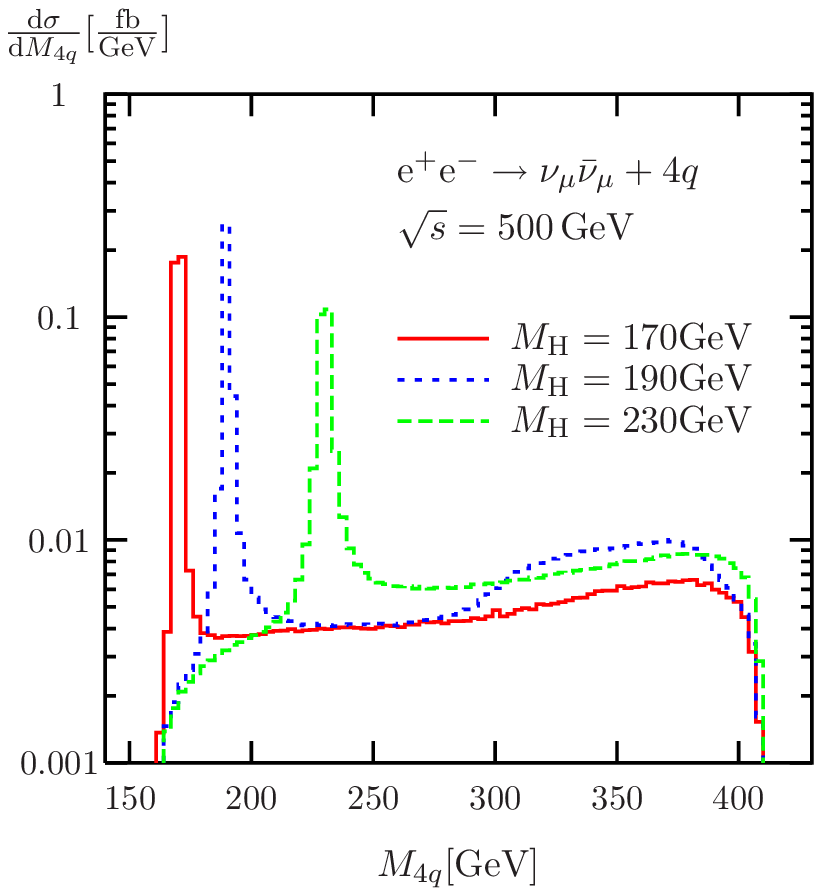}}
\put( 2.3,- 6.4){\includegraphics{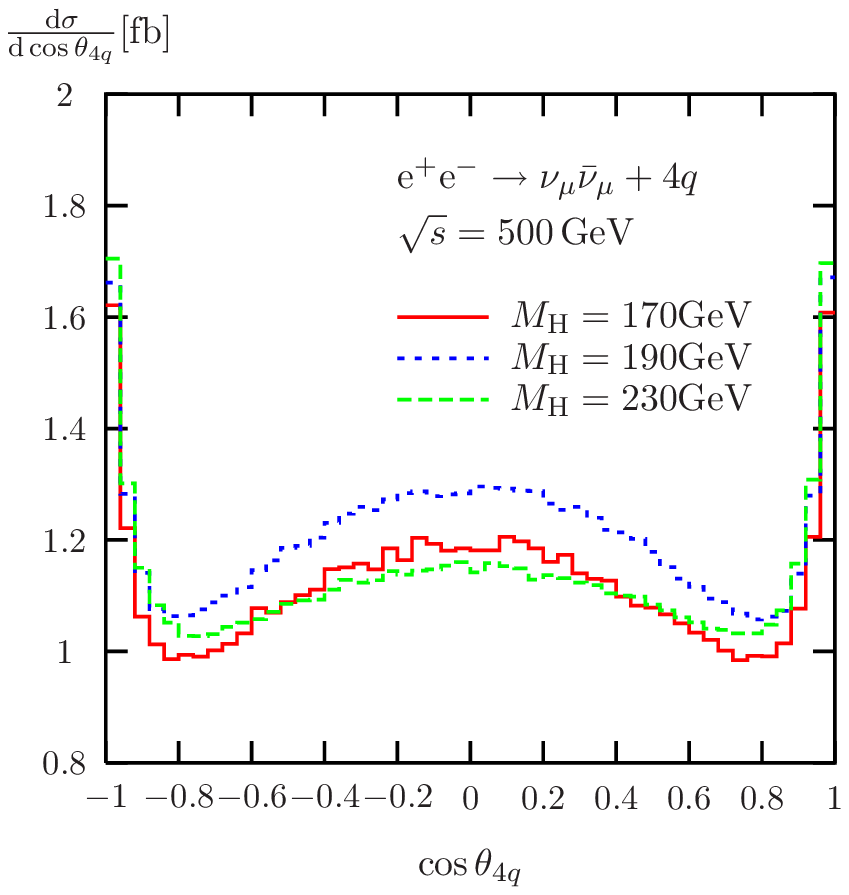}}
\put(-5.4,-12.9){\includegraphics{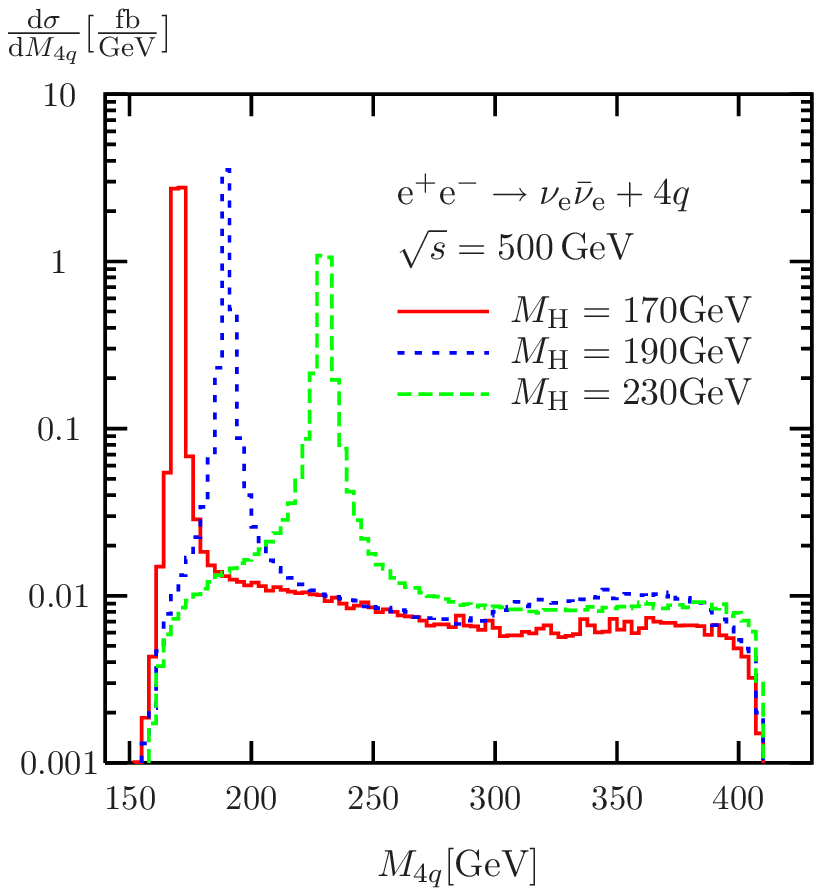}}
\put( 2.3,-12.9){\includegraphics{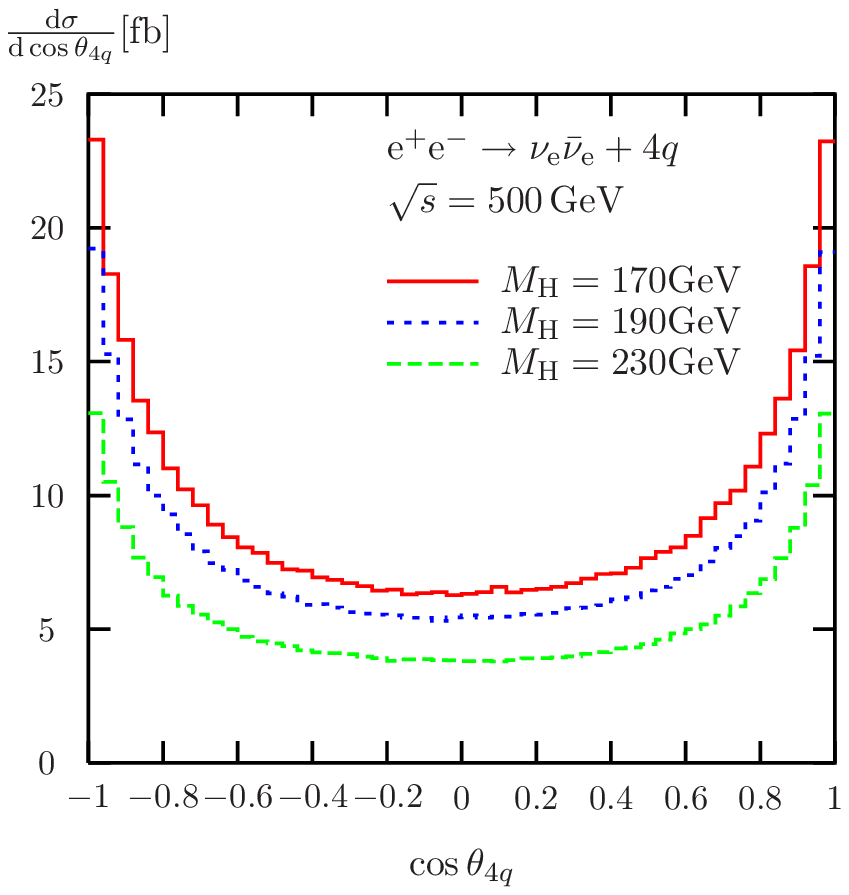}}
\end{picture} } 
\caption{Invariant-mass and angular distributions of the $4q$ system
in $\Pep\Pem\to(\nu_\mu\bar\nu_\mu/\nu_\Pe\bar\nu_\Pe)+4q$ 
(without ISR and gluon-exchange diagrams) for various Higgs masses}
\label{fig:higgs-nn4q}
\efi
\begin{table}[p]
\centerline{
\tabcolsep 3pt
\begin{tabular}{|c||c|c|c|c|c|}
\hline
$\sqrt{s}[\GeV]$ & 500 & 800 & 1000 & 2000 & 10000
\\ \hline \hline
fixed width & 1.6326(12) & 4.1046(35) &  5.6795(61) & 11.736(16) & 26.380(55)
\\ \hline
running width & 1.6398(12) & 4.1324(39) & 5.7206(54) & 12.881(14) & 12965(12)
\\ \hline
complex mass & 1.6330(12) & 4.1037(34) & 5.6705(54) & 11.730(14) & 26.387(57)
\\ \hline 
\end{tabular} }
\caption{Born cross sections in fb (without ISR) for 
$\Pep\Pem\to\nu_\Pe\bar\nu_\Pe\mu^-\bar\nu_\mu\Pu\bar\Pd$
for various CM energies and schemes for introducing decay widths}
\label{tab:wwww_width}
\end{table}
The crucial difference between the $\nu_\mu\bar\nu_\mu$ and 
$\nu_\Pe\bar\nu_\Pe$ channels lies in the Higgs production mechanisms:
while the former receives only contributions from $\PZ\PH$ production,
$\Pep\Pem\to\PZ+(\PH\to\PW\PW)\to 6f$,
the latter additionally involves W~fusion,
$\Pep\Pem\to\nu_\Pe\bar\nu_\Pe+(\PW\PW\to\PH\to\PW\PW)\to 6f$, 
which dominates the cross section.
Therefore, the cross section of $\nu_\Pe\bar\nu_\Pe+4q$ is an order
of magnitude larger than the one of $\nu_\mu\bar\nu_\mu+4q$.
The invariant-mass distributions of the two channels look similar,
both showing the resonance peaks for the decays $\PH\to\PW\PW\to 4q$
at $M_{4q}=\MH$ which appear over a continuous background.
Note that for $\MH=190\GeV$ and
$230\GeV$ the high-energy tails of the distributions show some
Higgs mass dependence. This is due to the subprocess of $\PZ\PH$
production where the Higgs decays into on-shell Z~bosons,
$\PH\to\PZ\PZ\to (\nu_\mu\bar\nu_\mu/\nu_\Pe\bar\nu_\Pe)+2q$,
which is not yet possible for the smaller Higgs mass $\MH=170\GeV$.
The corresponding boundary in $M_{4q}$,
which is clearly seen in the plots for $\MH=190\GeV$, 
is determined by the two extreme situations where the decay $\PH\to\PZ\PZ$
proceeds along the $\PZ\PH$ production axis. 
For $\MH=230\GeV$ this boundary is hidden by the Higgs peak and the
upper kinematical limit in the $M_{4q}$ spectrum.
In contrast to the
invariant-mass distributions, the shapes of the $4q$ angular
distributions of the $\nu_\mu\bar\nu_\mu$ and $\nu_\Pe\bar\nu_\Pe$
channels look very different. For $\nu_\mu\bar\nu_\mu$, i.e.\
for $\PZ\PH$ production, intermediate
production angles dominate, and this dominance is more pronounced
for smaller Higgs-boson masses, where more phase space is available.
For $\nu_\Pe\bar\nu_\Pe$, i.e.\ W-boson fusion, forward and backward 
production of Higgs bosons is preferred, and the $\MH$ dependence is 
mainly visible in the overall scale of the distribution, but not 
in the shape itself.

\section{Results On Vector-Boson Scattering}

Finally, in \refta{tab:wwww_width} 
we consider the high-energy behaviour of a typical channel involving
the subprocess $\PW\PW\to\PW\PW$, using different 
schemes for introducing finite decay widths.
This comparison is particularly important in order to control
gauge-invariance violating effects in several schemes.
In the {\it fixed-width scheme} all massive boson propagators
receive a constant width $\Gamma_B$ ($B=\PH,\PW,\PZ$), 
while in the {\it running width scheme} 
$\Gamma_B$ is multiplied by $p^2/M_B^2\times\theta(p^2)$, with $p^2$
denoting the virtuality of the propagator.
Both schemes violate gauge invariance.
In the {\it complex-mass scheme} \cite{Denner:1999gp}, gauge invariance
is restored by consistently using complex masses for the unstable
particles in the Feynman rules, i.e.\ it makes use of the propagators
of the fixed-width scheme and appropriately chosen complex couplings.
The example confirms the expectation from $4f(+\gamma)$ studies
\cite{Denner:1999gp}
that the fixed-width scheme, in spite of 
violating gauge invariance, practically yields the same results as
the complex-mass scheme.
In contrast, the running-width scheme breaks gauge invariance so badly that
deviations from the complex-mass scheme are already visible below $1\TeV$.
Above $1\TeV$ these deviations grow rapidly, and the high-energy limit
of the prediction is totally wrong.
Thus, if finite decay widths are introduced 
on cost of gauge invariance, the result is only reliable
if it has been compared to a gauge-invariant calculation, as it is for
instance provided by the complex-mass scheme. 
Moreover, our 
numerical studies (see also \citere{Dittmaier:2002ap})
show that the fixed-width scheme is in fact a good 
candidate for reliable results also in six-fermion 
production, although it does not respect 
gauge invariance. Whether this observation generalizes to
all $6f$ final states (or even further) is, however, not clear.

\end{document}